\documentclass[prb,aps,twocolumn,footinbib,floatfix,10pt,superscriptaddress]{revtex4-1}

\usepackage{chemformula} % Formula subscripts using \ch{}
\usepackage[T1]{fontenc} % Use modern font encodings
\usepackage{color}
\usepackage{graphicx}

\begin{document} %APS

\author{Patrick Laferrière}
\affiliation{National Research Council of Canada, Ottawa, Ontario, Canada, K1A 0R6.}
\affiliation{University of Ottawa, Ontario, Canada}
\author{Sofiane Haffouz}
\affiliation{National Research Council of Canada, Ottawa, Ontario, Canada, K1A 0R6.}
%\author{Edith Yeung}
%\altaffiliation{University of Ottawa, Ontario, Canada}
\author{David B. Northeast}
\affiliation{National Research Council of Canada, Ottawa, Ontario, Canada, K1A 0R6.}
\author{Philip J. Poole}
\affiliation{National Research Council of Canada, Ottawa, Ontario, Canada, K1A 0R6.}
\author{Robin L. Williams}
\affiliation{National Research Council of Canada, Ottawa, Ontario, Canada, K1A 0R6.}
\author{Dan Dalacu}
\affiliation{National Research Council of Canada, Ottawa, Ontario, Canada, K1A 0R6.}
\affiliation{University of Ottawa, Ontario, Canada}

\title{Position-controlled Telecom Single Photon Emitters Operating at Elevated Temperatures}

%\begin{document} %ACS

\preprint{APS/123-QED}


%\begin{tocentry}

%\includegraphics*[width=8.2cm,clip=true]{TOC.eps}
 
%\end{tocentry}


\begin{abstract}
Single photon emitters are a key component for enabling the practical use of quantum key distribution protocols for secure communications. For long-haul optical networks it is imperative to use photons at wavelengths that are compatible with standard single mode fibers: 1.31\,$\mu$m and 1.55\,$\mu$m. We demonstrate high purity single photon emission at $1.31\,\mu$m using deterministically positioned InP photonic waveguide nanowires containing single InAsP quantum dot-in-a-rod structures. At 4\,K the detected count rate in fiber was 1.9\,Mcps under above-band pulsed laser excitation at 80\,MHz corresponding to a single photon collection efficiency at the first lens of 25\%. At this count rate, the probability of multiphoton emission is $g^{(2)}(0)=0.021$. We have also evaluated the performance of the source as a function of temperature. Multiphoton emission probability increases with temperature with values of 0.11, 0.34 and 0.57 at 77\,K, 220\,K and 300\,K, respectively, which is attributed to an overlap of temperature-broadened excitonic emission lines. These results are a promising step towards scalably fabricating telecom single photon emitters that operate under relaxed cooling requirements.

\end{abstract}

\maketitle %APS
%\newpage

%$Keywords: quantum dot, nanowire, photonic waveguide, selective-area vapor-liquid-solid epitaxy, single photon source.
%\vspace{0.5cm}

The prospect of a future fiber-based quantum network \cite{Kimble_NAT2008} relies on the on-demand availability of high quality single photons at telecom wavelengths. Sources based on semiconductor quantum dots, which can be tuned over a wide wavelength range through the choice of material system \cite{Arakawa_APR2020}, are a promising candidate for delivering such photons. Telecom single photon emission was first demonstrated in both the InAs/GaAs \cite{Ward_APL2005} and the InAs/InP \cite{Miyazawa_JJAP2005} material systems in 2005. Since these early studies, device performance has dramatically improved for both GaAs-based \cite{Zinoni_APL2006,Xue_APL2017,Chen_NRL2017,Paul_APL2017,Carmesin_2018,Srocka_APL2020,Holewa_SR2020,Bauer_APL2021,Nawrath_APL2021,Zeuner_ACSP2021,Kolatschek_NL2021} and InP-based \cite{Takemoto_JAP2007,Birowosuto_SR2012,Benyoucef_APL2013,Takemoto_SR2015,Miyazawa_APL2016,Kim_Optica2016,Dusanowski_APL2016,Kim_NL2017,Haffouz_NL2018,Muller_NC2018,Jaffar_NS2019,Lee_APL2019,Anderson_QI2020,Lee_NL2021,Musial_APL2021,Holewa_ACSP2022} quantum dots with current state-of-the-art devices possessing a first-lens brightness of 36\% in the InP system using photonic crystal cavities \cite{Kim_Optica2016} and 23\% in the GaAs system using circular Bragg gratings \cite{Kolatschek_NL2021}.

\begin{figure}[htb]
\begin{center}
\includegraphics*[width=8cm,clip=true]{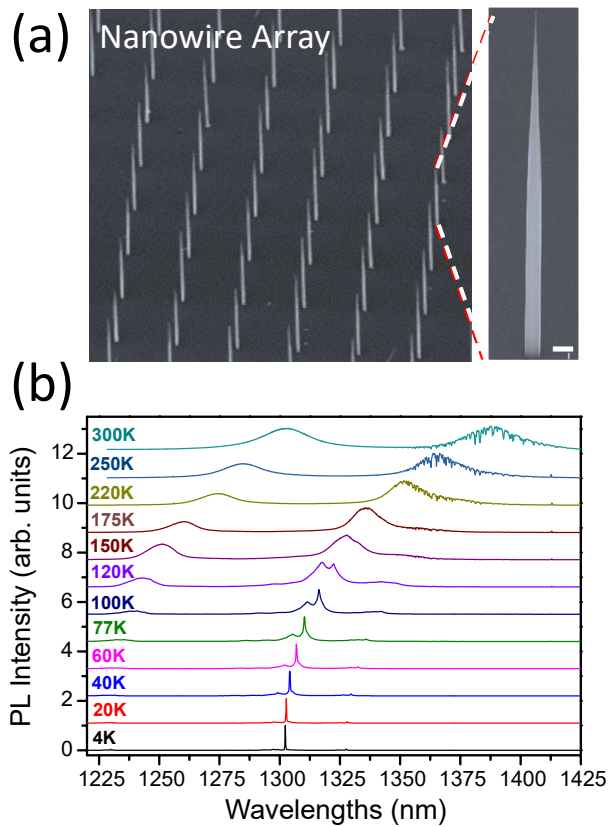}
\end{center}
\caption{(a) Scanning electron microscopy image at $45^\circ$ of an array of nanowires pitched at $7.5\,\mu$m. Right panel shows a zoom in of a single nanowire where the scale bar is 300\, nm. (b) PL spectra as a function of temperature of a single nanowire pumped at P$_\mathrm{sat}$ from 4\,K to 300\,K.}\label{fig1}
\end{figure}

Here we report state-of-the-art O-band quantum dot single photon emitters manufactured using a position-control technique that can operate at elevated temperatures. The emitters are based on bottom-up InAsP/InP nanowire quantum dots embedded within photonic nanowire waveguides grown using selective-area vapour-liquid-solid (S-A VLS) epitaxy \cite{Dalacu_APL2011}. In an earlier study \cite{Haffouz_NL2018} we tuned the quantum dot emission to telecom wavelengths by increasing the dot thickness, since memory effects of the Group V concentration in the gold catalyst limited the amount of arsenic that could be incorporated. Device collection efficiencies were low, which we attribute to a ground state consisting of both heavy- and light-hole levels \cite{Zielinski_PRB2013,Jeannin_PRB2017}, the latter of which results in photons that do not couple to the fundamental mode of the nanowire waveguide \cite{Jeannin_PRApp2017}. Here we adopt a dot-in-a-rod (DROD) structure \cite{Haffouz_APL2020} such that the thickness of the dot is nominally the same as in devices which have demonstrated high collection efficiencies \cite{Laferriere_SR2022}, and further optimised the nanowire geometry for O-band emission.

At 4\,K, the devices generated 1.9\,Mcps at the detector when excited at a rate of 80\,MHz, corresponding to an end-to-end efficiency of 2.6\% and a first lens single photon collection efficiency of 25\%. At these count rates, the probability of multiphoton emission was $2.1\%$. The devices also demonstrated single photon emission up to 220\,K, albeit at significantly reduced single photon purities of $\sim 34\%$, which permits operation with a simple multistage thermoelectric cooler.

The nanowires were grown by chemical beam epitaxy using the S-A VLS growth technique that allows positioning of individual emitters at specified locations on the growth substrate, see Refs.~\citenum{Dalacu_APL2011} and \citenum{Haffouz_APL2020} for details. A single InAs$_x$P$_{1-x}$ quantum dot with composition $x \sim 0.68$, diameter $D \sim 20$\,nm and thickness $t_d \sim 3$\,nm was embedded within an InAs$_y$P$_{1-y}$ nanowire rod with composition $y \sim 0.5$, diameter $D \sim 20$\,nm and thickness $t_r \sim 20$\,nm, itself embedded within an InP nanowire core. The core was clad with an InP shell to produce a photonic waveguide targeting a base diameter of 310\,nm tapered to 20\,nm over the 12$\,\mu$m length of the waveguide, see Fig.~\ref{fig1}(a).

Temperature-dependent optical measurements were made with the device in a closed-cycle helium cryostat. Above-band pulsed excitation at $\lambda = 670$\,nm was provided through a 100x cryogenic objective (numerical aperture NA = 0.81) located in the cryostat. Emission was collected through the same objective and directed to a spectrometer with a liquid nitrogen-cooled InGaAs detector for spectrally resolved measurements. For time-resolved measurements emission was coupled to a fiber and sent to two superconducting nanowire single photon detectors (SNSPDs) via a 50/50 beamsplitter. For the latter case the emission line to be measured was isolated using a filter with a bandwidth appropriate for the measurement, discussed below. 

\begin{figure*}%[h]
\begin{center}
\includegraphics*[width=15cm,clip=true]{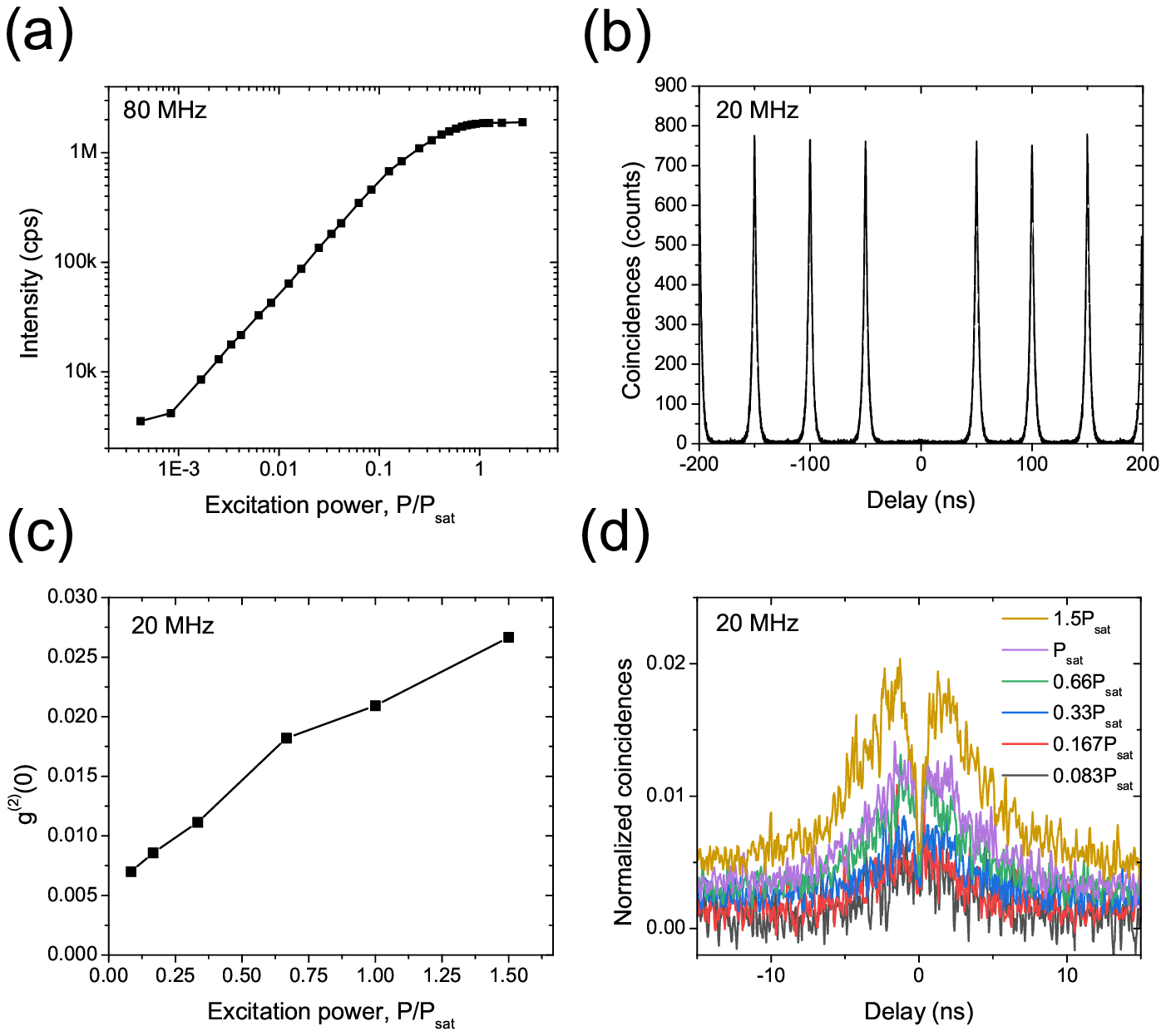}
\end{center}
\caption{Low temperature (4\,K) measurements: (a) Detected counts as a function of excitation power using an 80\,MHz pulsed source showing a saturation count level of 1.9\,Mcps. (b) Coincidence counts measured at $0.1\,P_\mathrm{sat}$ pumped at 20\,MHz for which g$^{(2)}(0)= 0.007$. (c) Integrated counts in the g$^{(2)}(\tau)$ zero-delay peak relative to side peaks as a function of excitation power at 20\,MHz. (d) Zoom in of the correlation peak around $\tau=0$\,ns as a function of excitation power showing evidence of re-excitation.}\label{fig2}
 \end{figure*}

Typical photoluminescence spectra as a function of temperature from a device emitting around 1300\,nm are shown in Fig.~\ref{fig1}(b) at an excitation power $P=P_\mathrm{sat}$ which corresponds to the power required to saturate the excitonic transition. At 4\,K the PL spectrum is characteristically dominated by a narrow single peak with a full width half maximum value of $45$\,$\mu$eV, limited by the resolution of the spectrometer. The peak is attributed to an excitonic transition within the quantum dot and, with increasing temperature, shifts continuously to longer wavelength, starting at 1301.28\,nm at 4K and reaching 1397.8\,nm at 300K. As the temperature is increased the predominant peak broadens and starts to overlap with the adjacent higher energy transition line that rises up, becoming one single broad peak above 150\,K. Above 100\,K a broad short wavelength peak associated with p-shell transitions can also be seen. It should be noted that emission from a single quantum dot can be readily observed at room temperature.

Exciting at 80\,MHz we measured, at saturation, 1.86\,Mcps on the SNSPDs (3.8\,Mcps under continuous wave excitation) after filtering with a 0.1\,nm filter, see Fig.~\ref{fig2}(a). The end-to-end source efficiency, which is defined by dividing the maximum collected count rates at the SNSPDs by the repetition rate of the pumping laser (80\,MHz), is thus 2.3\% at saturation. If we account for the detector efficiency of 90\%, the end-to-end efficiency is 2.6\% and by taking into account a throughput of 10\% for the experimental setup a collection efficiency at the first-lens of 26\% is obtained. This value is reduced to 25\% after correcting for the multiphoton emission probability measured at the above count rate (see below). The efficiency increases to 32\% if we include photons emitted into the phonon sidebands, estimated to be 20\% of the total emission at 4\,K \cite{Lodahl_RMP2015}, which we have filtered out using the narrow passband filter. In principle the efficiency can be nearly doubled using a mirror (Au layer) at the bottom of the nanowire to allow collection of the photons that are emitted towards the InP substrate. A theoretical model reflectivity of about 0.91 can be achieved using a gold layer under the nanowire base\cite{Claudon_NP2010}.

To assess the single-photon purity, second-order correlation measurements, g$^{(2)}$($\tau$), were performed in a standard Hanbury-Brown and Twiss experiment. Here, a pulse repetition rate of 20\,MHz was used to avoid significant overlap between consecutive pulses and a 0.1\,nm bandpass filter was used to isolate the emission from a single transition. A typical correlation measurement is shown in Fig.~\ref{fig2}(b) when the source is excited at $\sim 0.1\,P_\mathrm{sat}$. The integrated counts around $\tau=0$\,ns relative to the side peaks gives a probability of multiphoton emission of 0.7\%. As the pump power is increased the multiphoton probability increases, as shown in Fig.~\ref{fig2}(c), with a value 2.1\% at saturation. The reason for this increase is the presence of re-excitation of the dot due to the use of above band excitation, with a characteristic dip in the centre of the peak observed around zero delay\cite{Laferriere_NL2020,Aichele_NJP2004}, see Fig.~\ref{fig2}(d). This behaviour is a consequence of an excess of carriers in the barrier material at high pump powers that can be captured by the dot after the first photon emission event has occurred, leading to the re-excitation of the dot. The rise time from 0\,ns is associated with the carrier capture and subsequent relaxation within the dot, whilst the decay time is associated with the lifetime of the exciton.

\begin{figure*}%[h]
\begin{center}
\includegraphics*[width=15cm,clip=true]{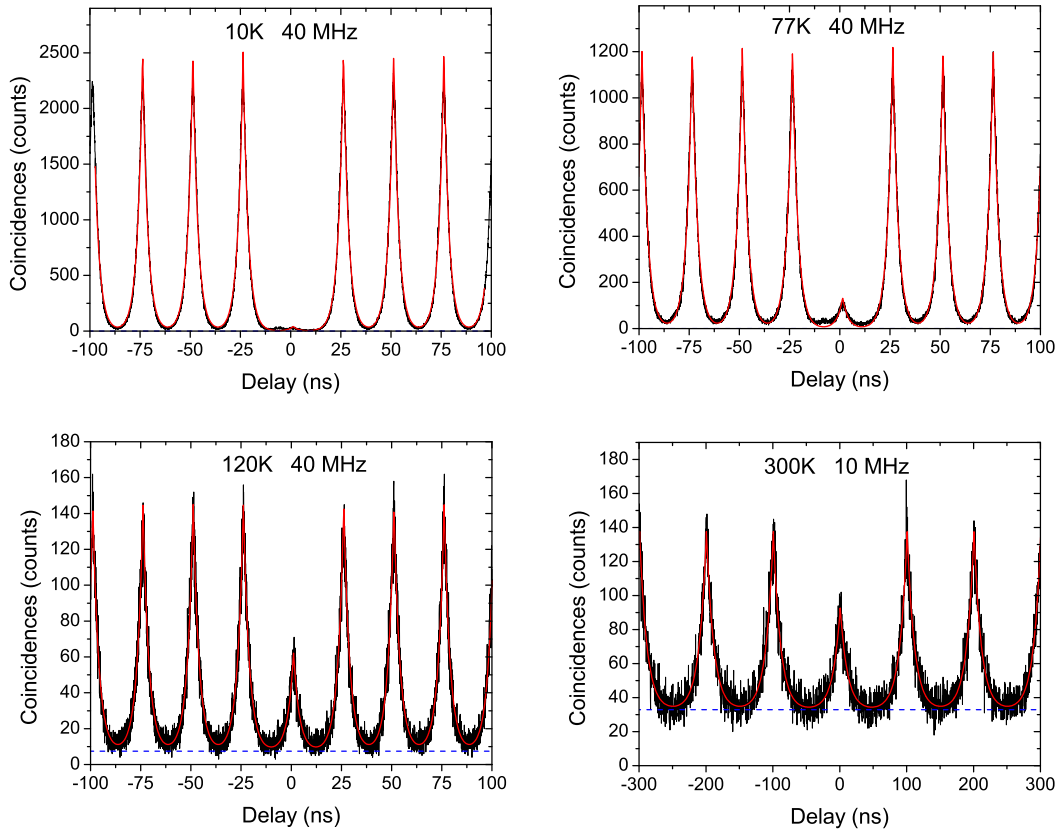}
\end{center}
\caption{Second-order correlation measurements of the source at different temperatures. The excitation power was 0.5\,P$_\mathrm{sat}$ at temperatures up to 120\,K and reduced to $\sim 0.25$\,P$_\mathrm{sat}$ at 220\,K and above. The red line is a curve fit to the data and the blue dashed line is the fitted background count level. A separate measurement of the lifetime was made to allow the background level to be determined.}\label{fig3}
 \end{figure*}

Next we assess the temperature dependence of the single-photon purity through coincidence measurements from 4\,K to 300\,K. In Fig.~\ref{fig3} we show correlations at four selected temperatures that span the measured range.  Before discussing the $g^{(2)}(\tau)$ dependence on temperature, we first note that as the temperature was increased the emission lines broadened and started to overlap, as shown in Fig.~\ref{fig1}(b). Importantly, the linewidth broadening with increasing temperature will have two effects on the second-order correlation measurements: First the use of a narrow pass band filter will reduce the count rates arriving at the SNSPDs. To maintain reasonable count rates for the measurements, the bandpass of the filter employed was widened as the temperature was raised, from 0.1\,nm to 12\,nm at 175\,K and to 25\,nm at 300\,K. Second, the overlap between the adjacent optical emission lines will make the selection of a single transition difficult. Both of the above can potentially result in photons from different transitions reaching the detectors and affecting the g$^{(2)}(0)$ value. If two transitions are not the result of a sequential decay in a cascade process, for example a neutral and charged exciton, then g$^{(2)}(0)$ can remain zero. On the other hand if the two transitions correspond to a cascade process, such as for the biexciton and exciton, then the value of g$^{(2)}(0)$ will rise.

To obtain accurate values of the coincidence counts in the zero delay peak, a curve fitting procedure was applied to the measured g$^{(2)}(\tau)$ curves which included an uncorrelated background signal. To enable accurate fitting a time-resolved photoluminescence (TRPL) measurement was made before each correlation measurement to determine the lifetime to be used for each fit. Without this extra piece of information a wide range of background levels and lifetimes could be used to fit the same data, in particular at high temperature. The results of the fitting are shown in Fig.~\ref{fig5} (red stars). 

\begin{figure}
\begin{center}
\includegraphics*[width=8cm,clip=true]{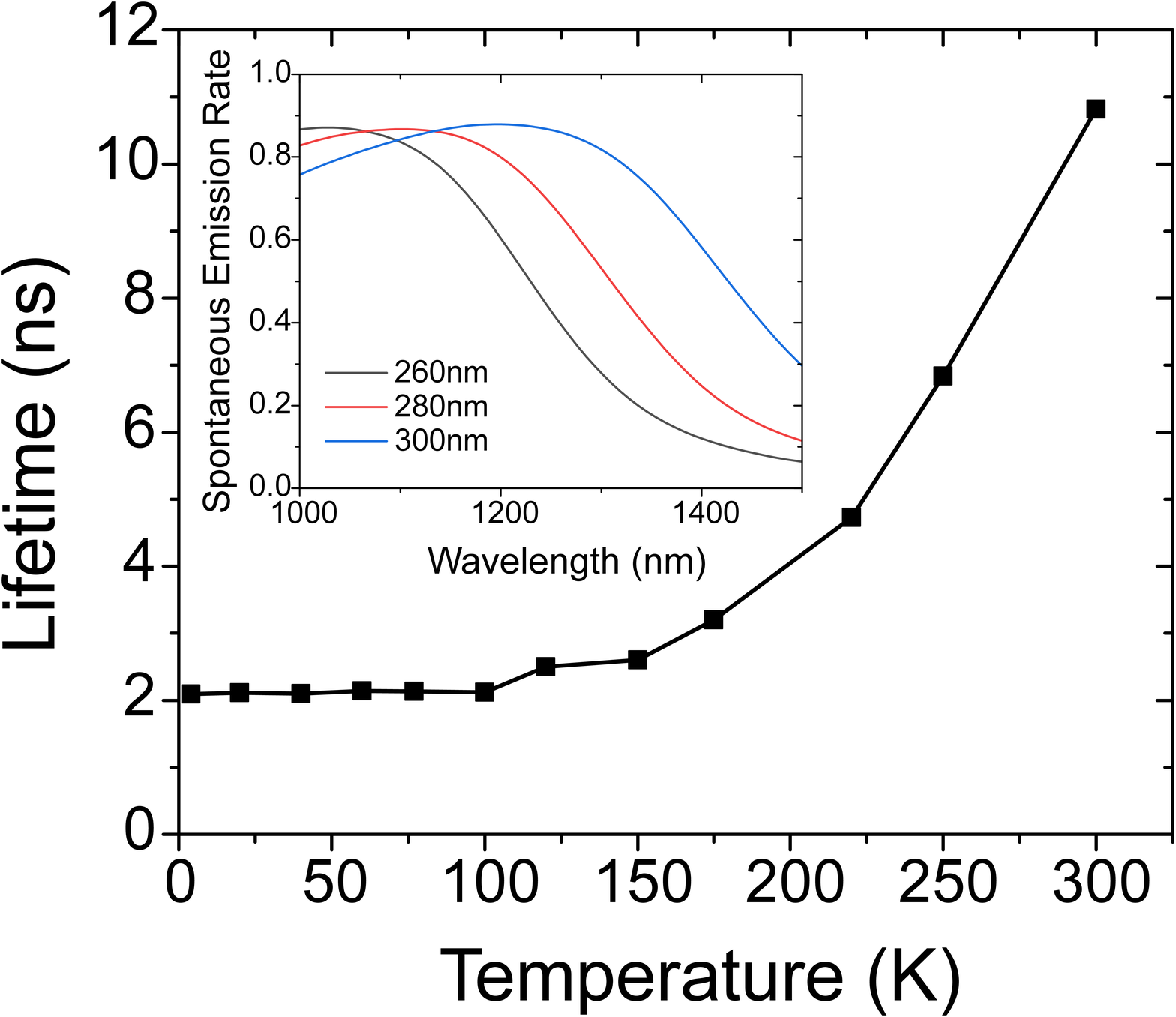}
\end{center}
\caption{Measured photoluminescence lifetime as a function of temperature. The inset shows the calculated spontaneous emission rate, normalised to that for bulk, as a function of wavelength for three different nanowire diameters.}\label{fig4}
\end{figure}

Before discussing the observed $g^{(2)}(0)$ dependence on temperature, we first look at the lifetime dependence extracted from the TRPL measurements, plotted in Fig.~\ref{fig4}. We observed a dramatic increase in lifetime  with increasing temperature, from 2.1\,ns at 4\,K to 10.8\,ns at 300\,K. We attribute this increase in part to a dependence of the spontaneous emission rate into the detected waveguide mode, HE$_{11}$, on the emission wavelength\cite{Haffouz_NL2018}. The inset in Fig.~\ref{fig4} shows the calculated spontaneous emission rate, normalised to that for bulk, as a function of wavelength for three different nanowire diameters. For the nanowire diameters used here the shift in emission wavelength from 1300 to 1400\,nm, as seen when going from 4\,K to 300\,K, results in a drop in emission rate or equivalently an increase in lifetime.

This alone is insufficient to account for the observed increase and we consider a second mechanism proposed in Ref.~\citenum{Harbord_PRB2009} to describe the increase in lifetime with temperature observed in InAs/GaAs quantum dots for temperatures up to 250\,K. The authors attributed the increase to Boltzmann spreading over dark states, i.e. thermal excitation of electrons and holes out of the ground state into higher lying states that do not have allowed optical transitions. Based on an estimate of the level spacing in our dot, from an s-p splitting of 68\,meV, this mechanism alone underestimates the increase in lifetime observed. If, however, we include both of the above mechanisms we can reach qualitative agreement with the observed lifetimes.

We consider now the temperature dependence of $g^{(2)}(0)$ plotted in Fig.~\ref{fig5} that shows a steady rise as the temperature is increased from 10\,K to $120\,$K. Above $120\,$K it saturates at just under 0.5 until 300\,K where it is just over 0.5. A measured non-zero g$^{(2)}(0)$ for a quantum dot can arise from a number of mechanisms; detector dark counts, scattered excitation laser, multiple independent emitters, re-excitation of the dot from the same excitation pulse, and multiple emission lines from a single dot. The increase in $g^{(2)}(0)$ with increasing temperature is clearly not a consequence of dark counts (which would be a uniform background), or scattered laser emission (which would be a narrower peak with a width limited by the detector timing jitter of 60\,ps). Multiple independent emitters are also very unlikely since the nanowire can contain one and only one dot by design and has a very low g$^{(2)}(0)$ at low temperature. This leaves re-excitation and multiple emission events from the dot as the likely causes. Re-excitation of the dot typically results in a dip in the g$^{(2)}(\tau)$ value around zero delay, which is not observed here, although the timescale for re-excitation at higher temperatures may be too short to resolve.

Multiple emission events from the dot, such as biexciton followed by exciton emission are also quite likely. At low temperature these transitions can be separated spectrally. With increasing temperature these transitions broaden and eventually overlap so they can no longer be isolated, increasing $g^{(2)}(0)$. By dropping the excitation power at high temperatures the g$^{(2)}(0)$ should improve due to the decreased probability of creating a biexciton, and indeed this is observed, but at the expense of overall count rates. Considering the apparent saturation $g^{(2)}(0)$ at a value of $\sim 0.5$ (i.e. two photons per excitation pulse) observed in Fig.~\ref{fig5}, it is likely that this last mechanism, involving only a single additional transition, dominates the observed temperature dependence. 

For comparison, we also plot literature values of $g^{(2)}(0)$ from other quantum dot-based single photon sources emitting at telecom wavelengths. Only results obtained from non-post-selected measurements (i.e. using pulsed excitation) are included. Measurements where the excitation power was specified to be at or close to $P_\mathrm{sat}$, as is the case here, are indicated by an asterisk in the legend. This applies strictly to the low temperature measurements since at higher temperatures it is difficult to observe saturation of a single transition when it overlaps with another. 

We first consider experiments performed using above-band excitation, as is the case here, indicated in the figure by filled symbols. The sources in this study, under these operating conditions, outperform, to our knowledge, all other existing approaches based on quantum dots. Next we consider experiments performed using quasi-resonant excitation (i.e. exciation via a p-shell level in the dot), indicated in the figure by open symbols. Here we observe several sources\cite{Miyazawa_APL2016,Takemoto_SR2015,Kolatschek_NL2021} that display reduced multiphoton emission probabilities compared to the devices in this study. In approaches utilizing randomly nucleated quantum dots such that multiple emitters may be simultaneous probed, quasi-resonant excitation will mitigate both spectral pollution from other emitters as well as re-excitation from the same emitter. For the sources studied here, which are fabricated using a site-selection technique that assures each device contains only one emitter, quasi-resonant excitation is also expected to improve the single photon purity by mitigating re-excitation process responsible for the non-zero $g^{(2)}(0)$ values, specifically at lower temperatures (see Fig.~\ref{fig2}(d)). 

\begin{figure}
\begin{center}
\includegraphics*[width=8cm,clip=true]{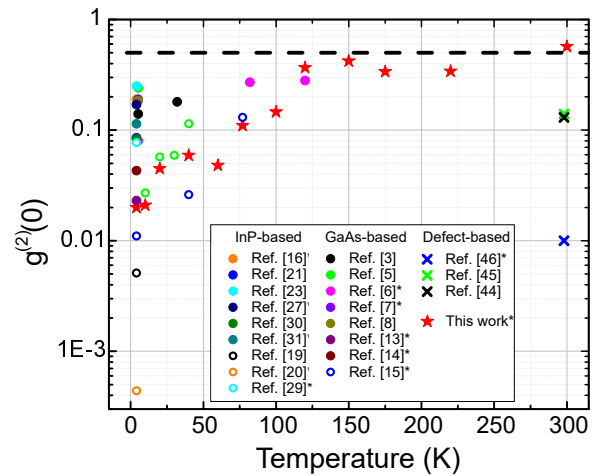}
\end{center}
\caption{Temperature dependence of $g^2(0)$ for quantum dot-based telecom single photon emitters. Dashed line corresponds to $g^2(0)=0.5$. Filled symbols: above-band excitation. Open symbols: p-shell excitation. Crosses: Defect-based sources. Values measured at close to maximum count rates indicated by an asterisk in the legend.}\label{fig5}
\end{figure}

For completion, we also include other solid-state 2-level systems that have demonstrated telecom single photon emission, specifically at room temperature, indicated in the figure by crosses. These include sources based on defects in SiC \cite{Wang_NC2018}, GaN \cite{Zhou_SciAdv2018} and carbon nanotubes \cite{He_NP2017}. Although the nanowire-based sources described here technically generate non-classical light up to temperatures of 220\,K, devices with such high multiphoton emission probabilities are of limited practical application \cite{Aharonovich_NP2016}. To achieve high temperature operation with reduced multiphoton emission probability to levels comparable to these defect-based systems will require engineering of the quantum dot electronic levels\cite{Cygorek_PRB2020} such that they are sufficiently separated when severely broadened to eliminate any overlap between them.  

In conclusion, we have demonstrated high purity telecom single photon emission from devices operating with high efficiency and grown using a position-control technique. We have also evaluated the temperature-dependent performance of the sources, quantifying the degradation of single photon purity with temperature. Finally, we note that the structures described can be incorporated in a hybrid on-chip platform\cite{Mnaymneh_AQT2020} to provide a stable and robust plug and play\cite{Northeast_SR2021} field-ready source that will be required to scalably build a future telecom quantum network.

%\begin{acknowledgement}

This work was supported by the Natural Sciences and Engineering Research Council of Canada through the Discovery Grant SNQLS, by the National Research Council of Canada through the Small Teams Ideation Program QPIC and by the Canadian Space Agency through a collaborative project entitled ‘Field Deployable Single Photon Emitters for Quantum Secured Communications'. 
%\end{acknowledgement}

%\begin{suppinfo}
%The following files are available free of charge.
%\begin{itemize}
%  \item Supporting Information: PL measurements of quantum dot emission linewidth versus temperature.
%  %\item Filename: brief description
%\end{itemize}
%\end{suppinfo}

\bibliography{whiskers}   
\end{document}